\documentstyle[12pt]{article}
\begin{document}

\baselineskip 6mm
\renewcommand{\thefootnote}{\fnsymbol{footnote}}

\newcommand{\nc}{\newcommand}
\newcommand{\rnc}{\renewcommand}

%\headheight=0truein
%\headsep=0truein
%\topmargin=0truein
%\oddsidemargin=0truein
%\evensidemargin=0truein
%\textheight=9truein
%\textwidth=6.5truein

\rnc{\baselinestretch}{1.24}    % 1.5 spacing btwn text lines
\setlength{\jot}{6pt}       % spacing btwn the rows of an eqnarray
\rnc{\arraystretch}{1.24}   % spacing btwn the rows of a non-eqn array

%%%%%%%%%%%%%%%%%%%%%% Equation Numbering %%%%%%%%%%%%%%%%%%%%%%%
%\makeatletter
%\rnc{\theequation}{\thesection.\arabic{equation}}
%\@addtoreset{equation}{section}
%\makeatother

%%%%%%%%%%%%%%%%%%%%%%%%%%%%%%%%%%%%%%%%%%%%%%%%%%%%%%%%%%%%%%%%%
%                                                               %
%                NEW COMMANDS AND MACROS                        %
%                                                               %
%%%%%%%%%%%%%%%%%%%%%%%%%%%%%%%%%%%%%%%%%%%%%%%%%%%%%%%%%%%%%%%%%

%%%%% Simplify some frequently used LaTeX commands %%%%%

\nc{\be}{\begin{equation}}

\nc{\ee}{\end{equation}}

\nc{\bea}{\begin{eqnarray}}

\nc{\eea}{\end{eqnarray}}

\nc{\ben}{\begin{eqnarray*}}

\nc{\een}{\end{eqnarray*}}

\nc{\xx}{\nonumber\\}

\nc{\ct}{\cite}

\nc{\la}{\label}

\nc{\eq}[1]{(\ref{#1})}

\nc{\newcaption}[1]{\centerline{\parbox{6in}{\caption{#1}}}}

\nc{\fig}[3]{

\begin{figure}
\centerline{\epsfxsize=#1\epsfbox{#2.eps}}
\newcaption{#3. \label{#2}}
\end{figure}
}

%%% Double line letters %%%

\def\IR{{\hbox{{\rm I}\kern-.2em\hbox{\rm R}}}}
\def\IB{{\hbox{{\rm I}\kern-.2em\hbox{\rm B}}}}
\def\IN{{\hbox{{\rm I}\kern-.2em\hbox{\rm N}}}}
\def\IC{\,\,{\hbox{{\rm I}\kern-.59em\hbox{\bf C}}}}
\def\IZ{{\hbox{{\rm Z}\kern-.4em\hbox{\rm Z}}}}
\def\IP{{\hbox{{\rm I}\kern-.2em\hbox{\rm P}}}}
\def\IH{{\hbox{{\rm I}\kern-.4em\hbox{\rm H}}}}
\def\ID{{\hbox{{\rm I}\kern-.2em\hbox{\rm D}}}}

%%%%% Roman pont in math

\def\Tr{{\rm Tr}\,}
\def\det{{\rm det}}

%%%%% Special Letters

\def\vare{\varepsilon}
\def\barz{\bar{z}}
\def\barw{\bar{w}}

\begin{titlepage}
%---------------- preprint number ---------------
\hfill\parbox{5cm} 
{IP/BBSR/2004-07 \\ hep-th/0403253}\\ 
\vspace{25mm}
\begin{center}
%------------------------ title ------------------------
{\Large {\bf Gravitational Waves in String theory\\
in Anti-de Sitter Background}  }

\vspace{15mm}
%---------------- authors and addresses ----------------
Alok Kumar\footnote{kumar@iopb.res.in}
\\[3mm]
{\sl Institute of Physics, Bhubaneswar 751 005, INDIA} \\
\end{center}

\thispagestyle{empty}

\vskip2cm

%----------------------- abstract ----------------------

\centerline{\bf ABSTRACT}
\vskip 4mm
\noindent

Inspired by the studies of gravitational waves in 
anti-de Sitter universe, in general relativity, in this paper
we investigate the possibility of similar solutions in 
IIB string theory 
on $AdS_3\times S^3 \times R^4$. We give a general form for such 
solutions in this background and present several explicit examples,
by directly solving the field equations, as well as the ones 
obtained by taking a scaling limit on  $D1-D5$ brane systems in a 
pp-wave background. The form of the metric in our solutions 
corresponds to a gravitational wave in $AdS_3$. We
show the supersymmetric  nature of these solutions
and discuss the possibility of their generalizations to 
other anti-de Sitter backgrounds, including the ones in four 
dimensions.
\\

\vspace{2cm}

\today

\end{titlepage}

Gravitational waves in general relativity, also known 
as pp-waves, have been a major research area for long 
time\cite{brinkmann,stefani}.
They provide a geometrical framework to understand the 
gravitational radiation, in addition to having possible
astrophysical relevance. However, space-times generally considered in 
these studies are asymptotically flat, the primary motivation being
the analysis of radiation from finite sources.  On the other hand,
exact solutions  representing gravitational wave in 
non-asymptotically flat backgrounds have also been a well-studied 
subject\cite{carmelli,podolsky1,podolsky2}. 
Such `cosmological' gravitational wave solutions are expected to provide 
a useful model for the propagation of primordial gravitational waves. 

Gravitational wave solutions in string theory
have also been a subject of great interest from various points of view. 
They provide exact solutions of string equations of 
motion\cite{amati-horowitz,tseytlin}  to all 
orders in inverse string tension, where the classical backgrounds 
do not receive worldsheet corrections. More recently,
an exactly solvable string construction has been presented in 
a special class of such ten dimensional string 
backgrounds\cite{blau,metsaev}, known as Hpp-waves, 
in the presence of constant flux for
the five-form  Ramon-Ramond (RR) field strength. 
Such configurations also appear in a `Penrose' limit of 
$AdS^5\times S^5$ string background\cite{penrose} 
and have been argued to be `dual' to 
four dimensional  gauge theories, with $N=4$ supersymmetry, 
in a sector with large R-charges{\cite{BMN,sheikh}. 
As a confirmation  of the conjecture, exact anomalous 
dimensions of gauge theory operators and various other 
predictions of string theory, have been 
verified using gauge theory techniques as well. This 
correspondence between string and gauge theories, 
also known as BMN duality, led to a surge of activity in 
the studies of plane-wave solutions in last few years. 
Our particular interest in this paper will be in the classical 
D-brane solutions in plane-wave backgrounds\cite{KNS-etc,biswas}.
Plane-wave solutions considered in string theory, including those
for BMN duality, however, have an interpretation 
of a wave structure in a flat Minkowski  background space-time. 

In this paper, we give, both the general  structure, as well as, 
explicit solutions for gravitational waves in an anti-de Sitter 
background in string theory. More precisely, we analyze 
ten-dimensional supergravity equations, following from 
IIB string theory, in an $AdS_3\times S^3\times R^4$
background, with a standard R-R 3-form flux. 
For a suitable ansatz, we find out the 
conditions under which one has a gravitational wave solution
in this background. Our result implies that, gravitational wave 
solutions can be generalized to string theory, 
by turning  on suitable components of $NS-NS$ 3-form fluxes, in addition
to the components of the metric, representing gravitational wave 
in $AdS_3$. Such components of metric and 
anti-symmetric tensor fields also appear in 
flat space pp-wave solutions. We then find out explicit 
examples of the gravitational waves in string theory, by first 
directly solving the conditions derived from the field equations in string 
theory. This solution has a wave structure only in the metric 
(since NS-NS three-form turns out to be absent in this case) and
the role of the additional field, namely R-R three-form is 
to compensate for the $AdS_3\times S^3$ curvature components.

We also obtain gravitational wave solutions by applying  
a scaling limit on the $D1-D5$ brane solution in a pp-wave 
background\cite{biswas}. 
The limiting solution satisfies the conditions
for the existence of the gravitational waves in string theory on 
$AdS_3\times S^3\times R^4$. The supersymmetric nature of this 
example is already guaranteed by the fact that the $D1-D5$ branes 
in the pp-wave background preserves a certain amount of supersymmetry. 
We, however, explicitly solve the Killing spinor equations
and show that the solutions presented in this paper are supersymmetric.
Several other gravitational wave solutions are also presented,
by using $S$ and $T$-duality transformations of string theory. 
We also discuss the possibility of the generalization of the results 
to four dimensional AdS space, by a suitable embedding 
in $M$-theory. 

We now begin by writing down the ansatz for the gravitational wave
solution in ten dimensional type IIB string theory on 
$AdS_3\times S^3 \times R^4$. First, to
write down the metric we use a structure which is 
similar to the one in the case
of pure gravity theory in four dimensions\cite{podolsky1}, 
known as Siklos  space-times:
\be
ds^2 = q \left\{ {du^2\over u^2} + {1\over u^2} (2 dx^+ dx^- 
       + H(u, x^+) {dx^+}^2) + d\Omega_3^2 \right\} + \sum_{i=1}^4 {dx^i}^2,  
\label{metric}
\ee
with `Poincare' coordinates $x^+, x^-$ and $u$ parameterizing $AdS_3$,
when  $H=0$. The function $H$ represents the gravitational wave profile. 
In our case, we assume the dependence of $H$ only on $u$, 
in addition to $x^+$. This implies that gravitational wave 
polarization lies in the $AdS_3$ part only. 
Many interesting examples in 
both flat  (such as Hpp-wave) as well as anti-de Sitter 
spaces  (as in the case of certain Siklos type solutions in four dimensional 
gravity\cite{podolsky1}) correspond to the case when $H$ is independent 
of $x^+$. One of our example below, following  from $D1-D5$ branes 
in a pp-wave 
background, also has a similar property. For the time 
being, however, we keep $H$ more general, 
dependent both on $u$ and $x^+$. In equation (\ref{metric}),
coordinates $x^i$ ($i=1,..,4$) represent $R^4$ 
coordinates and the metric on $S^3$ is parameterized as:
\be
   d\Omega_3^2 = d\theta^2 + cos^2\theta d\phi^2 + 
                 sin^2\theta d\psi^2. 
\label{deomega3}
\ee
In the following, indices $a, b$ ($a = 1, 2, 3$) 
etc. will also be used for denoting the coordinates on $S^3$. 
Full ten dimensional coordinates are denoted by Greek 
indices $\mu, \nu$ etc..
The $R^4$ part of the metric does not play any crucial role in our 
discussion, however they are needed to saturate the critical dimension
of the IIB string theory. Finally, the constant parameter $q$ 
in equation (\ref{metric}) gives the size or radius of curvature 
for $AdS_3$ and $S^3$ spaces.

One also recalls that in string theory, the $AdS_3\times S^3$ solutions
are accompanied by appropriate NS-NS or R-R 3-form
field strengths. To start
with, we take these as the later (R-R) possibility.  The R-R 3-form
field strengths, in the coordinates that we are working, have the 
following form: 
\be
F^{(3)} = \pm {2q \over u^3}dx^+ \wedge dx^- \wedge du 
          \pm 2 q \;\;sin\theta \;\;cos\theta \;\;
            d\theta \wedge d\phi \wedge d\psi \; .
\label{rr3-form}  
\ee
We notice the presence of $\pm$ signs in front of both the terms in 
the above equation. They correspond to the fact
that such a solution may originate  from either $D1$ 
($D5$)or $\bar{D1}$ ($\bar{D5}$)
branes. Equations of motion\cite{oz} 
are satisfied by all the 
possibilities, however, supersymmetry (in our notations below)
implies a relative sign between these terms. 
 
As is known, for $H=0$, the metric and 3-form field strengths 
given in equations (\ref{metric}) and (\ref{rr3-form}) already 
satisfy the supergravity field equations following from the 
type IIB string theory. For $H (u, x^+) \neq 0 $ one needs to 
turn on other fields as well. The situation is in fact 
similar to that in flat space, where for example, the associated
field strengths for the 3-form NS-NS fields are generally 
non-vanishing in directions transverse to the light-cone.
In the case of flat space, a suitable choice\cite{tseytlin} makes use of 
the field configurations, so that  dilaton ($\phi$) and 
NS-NS 3-form field strength $H^{(3)}$ take a form:
$\phi \equiv \phi (x^+)$, $H^{(3)}_{+IJ} = A_{IJ}(x^+)$, with $I, J$
running over the transverse directions.
With certain constraint, these fields, together with the 
flat space pp-wave metric, satisfy the field equations.
  
Guided by the general form of the NS-NS 3-form field strengths in the 
previous paragraph, for flat space pp-wave, we now propose the following 
form for them in the $AdS_3\times S^3\times R^4$ background:
\be
  H^{(3)} = dx^+ \wedge \left( ({q^2 A_a (x^a, x^+) \over u^3}) du 
            \wedge d x^a + ({q^2 B_{a b} (x^a, x^+) \over u^2}) 
             dx^a \wedge dx^b \right).  
\label{nsns3-form}
\ee
The form of $H^{(3)}$ in our case corresponds to the situation when 
(as in flat space ) non-zero components have been turned on 
for $H_{+\mu \nu}$'s  along all the directions of 
$AdS_3\times S^3$ that are
transverse to the light-cone. However, now there
is a non-trivial dependence of the components 
of $H_{+ \mu \nu}$ on the transverse 
coordinates as well. In equation (\ref{nsns3-form}), 
we have already identified the dependence on one of the coordinates 
transverse to the light-cone, namely $u$. The dependence of 
$A_a$'s and $B_{a b}$'s on $S^3$ coordinates will be 
determined later on from the field equations. The powers of 
$u$ in the coefficients of $A_a$ and $B_{a b}$ 
in equation (\ref{nsns3-form}) are also dictated by the field
equations which will be discussed below. 

For the gravitational wave 
solution that we are discussing, the metric in equation 
(\ref{metric}), R-R 3-form field strengths in equation 
(\ref{rr3-form}) and NS-NS 3-form field strengths in 
equation (\ref{nsns3-form}) are the only non-vanishing 
fields. All other fields of IIB theory, including the dilaton
are taken to be zero. Further restriction on the dilaton field
in the present case, unlike the one in flat space as mentioned above,
comes from a dilaton dependent factor in front of the 
contribution of the R-R fields in the gravitational field
equation of the IIB theory. 

The non-zero components of the 
Ricci tensor ($R_{\mu \nu}$) along directions $x^{\pm}, u$,  
for our choice of the metric in equation (\ref{metric}), are:
\be
   R_{++} = - {H_{, uu}\over 2} + {1\over 2}{H_{, u}\over u}
            - {2H\over u^2}, \;\;
   R_{+-} = - {2\over u^2}, \;\; R_{uu} = - {2\over u^2}\; .
\label{ricci-ads3}
\ee  
Other non-zero components of the Ricci tensor are along 
$S^3$ directions: 
\be
    R_{\theta \theta} = 2,\;\;\;R_{\phi \phi} = 2 cos^2\theta,\;\;\;
    R_{\psi \psi} = 2 sin^2 \theta .
\label{ricci-s3}
\ee

All the type IIB field equations are satisfied for our general 
ansatz in equations (\ref{metric}), (\ref{rr3-form}) and 
(\ref{nsns3-form}), provided a set of conditions are satisfied 
by functions $H(u, x^+)$, $A_a$'s and $B_{a b}$'s. The condition 
following from the field equation of the 
metric component $g_{+ +}$ is:
\be
    - {H_{, uu}\over 2} + {1\over 2} {H_{, u}\over u} 
         =  {q^2\over u^4} \left( {1\over 2} A_a A^a + 
           {1\over 4}  B_{a b} B^{a b} \right), 
\label{h-condition}
\ee
where the raising and lowering of the indices in $A_a$ and 
$B_{a b}$ has been done with respect to the metric 
on $S^3$ \footnote{Analogous condition for the pp-wave solution 
in flat space, mentioned earlier, has a form\cite{tseytlin}:
\be
\partial^I \partial_I H(x^+, x^I) + A_{IJ}A^{IJ} + 
                  4 \partial_{x^+}^2 \phi (x^+) = 0
\ee}. The similarity of the above condition (\ref{h-condition})
can also be seen with anti-de Sitter gravitational waves in 
four dimensional gravity\cite{podolsky1}, 
by setting  $A_a$'s and $B_{ab}$'s to zero.

The field equations for 3-form fields also imply the following 
conditions on quantities $A_a$'s and $B_{a b}$'s:
\be
\nabla^a A_a =0,\;\;\;A_a + {1\over 2} \nabla^b B_{a b} = 0, 
\label{AB-condition}
\ee 
with covariant derivatives being defined by the metric on $S^3$
given in equation (\ref{deomega3}). 
Finally, the Bianchi identity of $H$ gives a restriction:
\be
    \partial_a A_b - \partial_b A_a = - 2 B_{a b}.
\label{bianchi}
\ee

We have therefore presented the general structure of  gravitational waves
in type IIB string theory in $AdS_3\times S^3\times R^4$ background. 
They are characterized by functions $H$, $A_a$ 
and $B_{a b}$ which satisfy conditions (\ref{h-condition}), 
(\ref{AB-condition}) and (\ref{bianchi}). 
We now give explicit solutions for these 
conditions for two different choices of $H$, $A_a$ and $B_{a b}$'s.

First, a solution can be obtained by considering $A_a$ = $B_{a b} =0$,
so that NS-NS three-form field $H^{(3)}$ is trivial. Then one  only has 
a non-trivial R-R three-form flux given by equation (\ref{rr3-form}),
in addition to the metric (\ref{metric}). Moreover, the 
R-R three-form flux is identical 
to the one in the case of $AdS_3\times S^3$ background. 
This choice of  $A_a$ and $B_{a b}$ already 
satisfies equations (\ref{AB-condition}) and (\ref{bianchi}). 
Equation (\ref{h-condition}) in this case implies a general wave profile 
given by 
\be
         H(u, x^+) = f(x^+) u^2 + g(x^+),
\label{simple-solution}
\ee
with $f$ and $g$ being functions of $x^+$ only. 
Later on we will also discuss the supersymmetry property of this solution.

We now write down another gravitational wave solution by taking a scaling 
limit on a known $D1-D5$ brane in a pp-wave background\cite{biswas}.
In this case, the gravitational wave profile turns out to be 
independent of $x^+$ (similar to the case of Hpp-waves appearing in the 
case of `BMN' duality), although further generalization of this 
solution to include $x^+$ dependence is possible and will also be discussed
below. 

In fact, it can be verified that the choice of $A_a$ components: 
\be 
   A_{\theta} = 0,\;\;A_{\phi} = 2\mu cos^2 \theta,\;\;
         A_{\psi} = 2\mu sin^2\theta,
\label{A-components}
\ee
and nonzero $B_{a b}$ components:
\be
B_{\theta \phi} = \mu sin2\theta,\;\;
B_{\theta \psi} = - \mu sin2\theta, 
\label{B-components}
\ee
satisfy equations  (\ref{AB-condition}) and (\ref{bianchi}). 
The parameter $\mu$ in 
the above equation characterizes the gravitational wave. When 
$\mu = 0$,  one reduces to $AdS_3\times S^3$. Finally, equation 
(\ref{h-condition}) is also satisfied for:
\be
      H = -{\mu^2 q^2 \over u^2}.
\label{H-final}
\ee

Equations (\ref{A-components}), (\ref{B-components}) 
and (\ref{H-final}), together with 
the metric and 3-form fields  in equations (\ref{metric}), 
(\ref{rr3-form}) and (\ref{nsns3-form}), give a
gravitational wave solution in 
type IIB string theory on $AdS_3\times S^3\times R^4$. We now show the
connection of this solution to the $D1-D5$ branes in a pp-wave background
in type IIB string theory. 

The $D1-D5$ brane solution, reducing to the above gravitational wave
on $AdS_3\times S^3\times R^4$ in a scaling limit,
is given as \cite{biswas}:
\begin{eqnarray}
ds^2&=&(f_1 f_5)^{-{1\over 2}}(2 dx^+dx^-
-\mu^2{\sum_{i=1}^{4}}x_i^2 (dx^+)^2) + \Big({f_1 \over f_5}\Big)^{1\over
  2}{\sum_{a=5}^{8}}(dx^a)^2 \cr
& \cr
&+&(f_1 f_5)^{1\over 2}{\sum_{i=1}^{4}}(dx_i)^2, \cr
& \cr
e^{2\phi}&=&f_1\over f_5, \cr
& \cr
H^{(3)}_{+ 1 2}&=&H^{(3)}_{+ 3 4} = 2\mu,\cr
& \cr
F^{(3)}_{+ - i}&=& \partial_{i}{f_1}^{-1},~~~~
F^{(3)}_{m n p} = \epsilon_{m n p l}
\partial_{l}f_5,
\label{d1-d5}
\end{eqnarray}  
where $f_1$ and $f_5$ are the Green functions, in common transverse 
directions  $x^1,..,x^4$, representing $D1$ and $D5$ branes respectively:
$f_1 = 1 + {q_1\over r^2}$, $f_5 = 1 + {q_5\over r^2}$  and $r$ is
the radial coordinate in the transverse direction. In the 
scaling limit one takes
(see for example\cite{papado}): $r\rightarrow 0$, in addition to setting 
$q_1 = q_5 = q$. It can now be verified that such a scaling limit, 
together with a coordinate change $r = {q\over u}$, 
gives the gravitational wave solution of 
equations (\ref{metric}), (\ref{rr3-form}) and (\ref{nsns3-form})
when the wave profile is chosen as in equations  
(\ref{A-components}), (\ref{B-components}) and (\ref{H-final}). 
We have therefore 
shown that $D1-D5$ brane solution in pp-wave background gives, 
in a scaling limit, a gravitational wave in string theory in  
$AdS_3\times S^3\times R^4$ background.

We now discuss the supersymmetry properties of our solutions, first 
for the specific wave profile in equations (\ref{A-components}),
(\ref{B-components}) and (\ref{H-final}). 
Since this specific solution is obtained, in a scaling
limit, from a known BPS solution which already preserves a fraction of 
supersymmetry, we expect it to be supersymmetric as well.
Nevertheless, an explicit analysis is carried out below to 
find out the Killing spinors.  

Killing spinor equations follow from the 
dilatino and gravitino variations of the ten dimensional type IIB
supergravity\cite{hassan}. 
To write these equations, we make use of the 
form of the $S^3$ metric given in equation (\ref{deomega3}). 
Then, in order to satisfy the dilatino variation conditions,  
$\delta \lambda_{\pm} = 0$, one needs to 
impose the following conditions on the ten-dimensional supersymmetry
parameters, given by Majorana-Weyl spinors $\epsilon_{\pm}$
with positive chirality, and $\pm$ denote the 
left and the right moving sectors of the IIB string theory. 
The conditions are:
\be
   \Gamma^{\hat{+}} \epsilon_{\pm} = 0,
\label{gamma+}
\ee
\be
   (1 \pm \Gamma^{\hat{+}\hat{-}\hat{u}\hat{\theta}\hat{\phi}\hat{\psi}})
   \epsilon_{\pm} = 0,
\label{gamma-all}
\ee
where $\Gamma^{\hat{+}}$, $\Gamma^{\hat{-}}$ etc. are the Dirac 
gamma matrices, with superscripts denoting the tangent space indices.  
The $\pm$ signs in equation (\ref{gamma-all}) correspond to the 
choice of the relative signature in equation (\ref{rr3-form}).

Variations of gravitino imply several conditions depending on the 
components $\delta \psi_{\mu}$. In order to satisfy them, and 
solve these equations, one needs to impose:
\be
   \Gamma^{\hat{u}\hat{\theta}\hat{\phi}\hat{\psi}} \epsilon_{\pm} 
                 = \epsilon_{\pm},
\label{gamma+-}
\ee
in addition to equations (\ref{gamma+}) and (\ref{gamma-all}). 
Now, the consistency between 
equations (\ref{gamma+}), (\ref{gamma-all}) and (\ref{gamma+-})
\footnote{where we have used the fact that, 
by using a choice of the metric: 
$\{\Gamma^{\hat{+}}, \Gamma^{\hat{-}}\} = 2 $, condition
$\Gamma^{\hat{+}}\epsilon_{\pm} = 0$ is equivalent to 
$\Gamma^{\hat{+}\hat{-}}\epsilon_{\pm} = \epsilon_{\pm}$.}
implies that one needs to pick up the lower sign in equation 
(\ref{gamma-all}), in order to preserve supersymmetry. 

To clarify further, equation (\ref{gamma-all}) is the 
$1/2$ supersymmetry condition for the $AdS_3\times S^3\times R^4$ 
background in the absence of a gravitational wave. The other two 
conditions, namely (\ref{gamma+}) and (\ref{gamma+-}) arise when gravitational
wave, with the particular profile mentioned above
in equations (\ref{A-components}) - (\ref{H-final}), is present. 
By using these two conditions, all the Killing spinor equations,
appearing as first order differential equations, 
reduce precisely to the case of pure $AdS_3\times S^3\times R^4$
background. Since these equations are
known to admit the maximal number of Killing spinors,
residual supersymmetry is obtained by analyzing 
the projection conditions: (\ref{gamma+}), (\ref{gamma-all})
and (\ref{gamma+-}) on their solutions.   

Explicitly, the Killing spinor equations reduce to the following 
simple forms by  imposing the  supersymmetry conditions coming 
from the presence of  the gravitational waves, namely
equations (\ref{gamma+}) and (\ref{gamma+-}). By 
defining $\epsilon = \epsilon_+ + \epsilon_-$,
$\tilde{\epsilon} = \epsilon_+ - \epsilon_-$ we get:
\be
 \partial_u \epsilon + {1\over 2 u} \Gamma^{\hat{+}\hat{-}}\epsilon = 0,
\;\;\partial_-\epsilon - 
{1\over u} \Gamma^{\hat{+}\hat{u}}\epsilon = 0,\;\;
\partial_+ \epsilon = 0,
\label{epsilon-1}
\ee
\be
 \partial_u \tilde{\epsilon} -
{1\over 2 u} \Gamma^{\hat{+}\hat{-}}\tilde{\epsilon} = 0,\;\;
\partial_- \tilde{\epsilon} = 0,
\;\;\partial_+\tilde{\epsilon} - 
{1\over u} \Gamma^{\hat{-}\hat{u}}\tilde{\epsilon} = 0,
\label{tepsilon-1}
\ee
and
\be
\partial_{\theta} \epsilon - 
{1\over 2} \Gamma^{\hat{\phi}\hat{\psi}}\epsilon = 0,\;\;
\partial_{\phi} \epsilon + {sin\theta\over 2} 
\Gamma^{\hat{\theta}\hat{\phi}}\epsilon 
+ {cos \theta\over 2}\Gamma^{\hat{\theta}\hat{\psi}}\epsilon= 0,\;\;
\partial_{\psi} \epsilon - {cos\theta\over 2} 
\Gamma^{\hat{\theta}\hat{\psi}}\epsilon 
- {sin \theta\over 2}\Gamma^{\hat{\theta}\hat{\phi}}\epsilon= 0,
\label{epsilon-2}
\ee
\be
\partial_{\theta} \tilde{\epsilon} +
{1\over 2} \Gamma^{\hat{\phi}\hat{\psi}}\epsilon = 0,\;\;
\partial_{\phi} \tilde{\epsilon} + {sin\theta\over 2} 
\Gamma^{\hat{\theta}\hat{\phi}}\tilde{\epsilon} 
- {cos \theta\over 2}\Gamma^{\hat{\theta}\hat{\psi}}
\tilde{\epsilon} = 0,\;\;
\partial_{\psi} \tilde{\epsilon} - {cos\theta\over 2} 
\Gamma^{\hat{\theta}\hat{\psi}}\tilde{\epsilon} 
+ {sin \theta\over 2}\Gamma^{\hat{\theta}\hat{\phi}}
\tilde{\epsilon}= 0.
\label{epsilon-3}
\ee

The set of above equations have simple solutions for both 
$\epsilon$ and $\tilde{\epsilon}$. They are:
\be
   \epsilon = (e^{{1\over 2}\Gamma^{\hat{\phi}\hat{\psi}}\theta}
              e^{- \Gamma^{\hat{\theta}\hat{\psi}}(\phi - \psi)})
              (e^{-{1\over 2}\Gamma^{\hat{+}\hat{-}}ln u}
              e^{\Gamma^{\hat{+}\hat{u}}x^-})\epsilon_0,
\label{final-epsilon}              
\ee
and
\be
   \tilde{\epsilon} = (e^{- {1\over 2}\Gamma^{\hat{\phi}\hat{\psi}}\theta}
              e^{ \Gamma^{\hat{\theta}\hat{\psi}}(\phi + \psi)})
              (e^{{1\over 2}\Gamma^{\hat{+}\hat{-}}ln u}
              e^{\Gamma^{\hat{-}\hat{u}}x^+})\tilde{\epsilon}_0,
\label{final-tepsilon}              
\ee
where $\epsilon_0$ and $\tilde{\epsilon}_0$ are constant spinors.

Now, to count the amount of supersymmetry that is preserved, 
we notice that the supersymmetry projection (\ref{gamma-all})
commutes with all the operators appearing in the exponents in  
expressions (\ref{final-epsilon})
and (\ref{final-tepsilon}). As a result, this projection
gives a condition identical to (\ref{gamma-all}), when acting on 
$\epsilon_0$ and $\tilde{\epsilon}_0$. This is the $1/2$ 
supersymmetry in the absence of the gravitational wave. 

The remaining projections, (\ref{gamma+}) and (\ref{gamma+-}) on 
spinor $\epsilon$ also lead to identical conditions on 
$\epsilon_0$. This, however, does not hold when considering the spinor 
$\tilde{\epsilon}$ in equation (\ref{final-tepsilon}) due to 
the presence of $\Gamma^{\hat{-}}$. One is therefore left with 
only a Killing spinor $\epsilon$ in the presence of 
the gravitational wave. This spinor is also further restricted by two 
independent projections (\ref{gamma+}) and 
(\ref{gamma-all}). Thus we finally have $1/8$ supersymmetry for our 
solution.

Several other gravitational wave solutions in $AdS_3\times S^3\times R^4$ 
background can  be constructed using `duality' symmetries of string theory.
The simplest ones of these comes from a direct application of 
S-duality on our general solution in equations 
(\ref{metric}), (\ref{rr3-form}) and (\ref{nsns3-form}). This symmetry 
leaves the metric unchanged, however, NS-NS and R-R 3-form fields are
interchanged among themselves under the symmetry transformation. 
The specific wave profiles of both the solutions mentioned in the 
paper, namely the ones in (\ref{simple-solution}) and 
(\ref{A-components})-(\ref{H-final}), remain unchanged under the
symmetry transformation. We do not write these solutions explicitly. 

A more non-trivial example involves several applications of 
$S$ and $T$-dualities, amounting to taking a scaling limit on 
another $D1-D5$ brane solution\cite{biswas}, involving 
now self-dual R-R 5-form 
field strengths $(F^{(5)})$ of the IIB string theory, rather then 
NS-NS 3-forms $H^{(3)}$ as in equation  (\ref{d1-d5}). 
The gravitational wave solution obtained by taking a scaling limit
has the identical metric and 3-form flux as in 
equations (\ref{metric}) and (\ref{rr3-form}). However, one now
has a (self-dual) R-R 5-form flux (in place of NS-NS 3-form of 
equation (\ref{nsns3-form})):
\be
  F^{(5)} = dx^+ \wedge \left( ({q^2 A_a (x^a, x^+) \over u^3}) du 
            \wedge d x^a + ({q^2 B_{a b} (x^a, x^+) \over u^2}) 
             dx^a \wedge dx^b \right)\wedge (dx^1\wedge dx^2 + 
             dx^3\wedge dx^4),  
\label{rr5-form}
\ee
and, as already mentioned earlier,  $x^1,..,x^4$ are the coordinates 
of $R^4$. The wave profile for this gravitational wave solution
is identical to the one in  (\ref{A-components}),
(\ref{B-components}) and (\ref{H-final}). 

Further generalization of the gravitational wave solution, as 
obtained above from the $D1-D5$ branes in equations 
(\ref{A-components}) - (\ref{H-final}),
can be incorporated in string theory.
It is evident from equations (\ref{h-condition}), (\ref{AB-condition})
and (\ref{bianchi}) that one can obtain a general class of 
explicit solutions 
by multiplying $H$ in equation (\ref{H-final}) by any function 
$F(x^+)$, provided $A_a$ and $B_{a b}$ in equations (\ref{A-components})
and (\ref{B-components}) are multiplied by another function $G(x^+)$,
satisfying $F(x^+) = {G(x^+)}^2$. Supersymmetry analysis for 
these $x^+$ dependent solutions proceeds in a similar fashion 
as presented above.  The amount of supersymmetry also turns out 
to be identical. The primary reason for the similarity in the 
supersymmetry analysis is the fact that the supersymmetry 
projections coming from the gravitational waves,
namely equations (\ref{gamma+}) and (\ref{gamma+-}), 
still remain valid and reduce the Killing spinor equations once 
again to the same ones as in equations (\ref{epsilon-1}) -
(\ref{epsilon-3}). One also has an identical amount of supersymmetry
for another $x^+$ dependent solution presented in equation 
(\ref{simple-solution}). The difference in the supersymmetry analysis,
in this case, is the 
absence of the projection (\ref{gamma+-}). However, since 
this condition is not an independent one, as discussed earlier,
the amount of supersymmetry remains same.

We have therefore presented explicit solutions in string theory
representing gravitational waves in $AdS_3$. It is possible
to generalize our results in various ways. 
First, it may be possible to use other branes in pp-wave backgrounds,
in order to obtain gravitational waves in various other 
anti-de Sitter spaces in string theory. In particular, in our view,
an  interesting possibility is to look for gravitational waves in $AdS_4$,
through a suitable solution in   
$AdS_4\times S^7$ background of M-theory. This can possibly be done 
by using an $M2$ brane solution in a pp-wave background, in the same
way as above. Alternatively, one may be able to solve 
the M-theory equations of motion
directly, with a suitable ansatz, like the one given in 
equations (\ref{metric}), (\ref{rr3-form}) and (\ref{nsns3-form})
for $AdS_3$. Such solutions will provide the generalization
of `Siklos space-times', discussed in pure gravity, to M-theory. 
Another interesting aspect will be to examine if any of the 
gravitational wave solutions presented here provide exact string 
solutions, like the pp-waves or some $AdS_3\times S^3$ backgrounds.
It will also be useful to find out the connection of our results to 
certain shock wave solutions found in \cite{lunin}. 

{\bf Note Added:} After the submission of this paper to the archive,
I have also come to know of some other papers\cite{brecher,patri} 
where gravitational waves in anti-de Sitter spaces have been discussed. 
However, these gravitational waves correspond, in our language, 
to the situation when $A_a =  B_{a b} = 0 $ in our 
equations (\ref{nsns3-form}) and (\ref{rr5-form}). 
In particular, our $A_a = B_{a b} = 0$ solution (\ref{simple-solution}) 
has been given earlier in \cite{brecher} and shown to be equivalent 
to the pure $AdS_3$ without a gravitational wave. However, it is 
emphasized that the general situation above represents 
new gravitational wave structure in anti-de Sitter backgrounds 
with a non-trivial NS-NS 3-form (or R-R 5-form)
flux mixing $AdS_3$ and $S^3$ spaces.

{\bf Acknowledgment:} I would like to thank, A. Biswas,
B. Chandrasekhar, S. Mukherji and R.R. Nayak  
for several useful discussions and for pointing out some important
references. I would also like to thank O. Lunin and D. Brecher for 
pointing out the results appearing in \cite{lunin,brecher}.

%%%%%%%%%%%%%%%%%%%%%%%%%%%%%%%%%%%%%%%%%%%%%%%%%%%%%%%%%%%%%%%%%%%%%%%%
%                       REFERENCES                                     %
%%%%%%%%%%%%%%%%%%%%%%%%%%%%%%%%%%%%%%%%%%%%%%%%%%%%%%%%%%%%%%%%%%%%%%%%
%\newpage

\renewcommand{\thefootnote}{\arabic{footnote}}
\setcounter{footnote}{0}

%\tableofcontents
%%%%%%%%%%%%%%%%%%%%%%%%%%%%%%%%%%%%%%%%%%%%%%%%%%%%%%%%%%%%%%%%%%%%%%
%\section{Introduction}
%%%%%%%%%%%%%%%%%%%%%%%%%%%%%%%%%%%%%%%%%%%%%%%%%%%%%%%%%%%%%%%%%%%%%%

\end{document}